\documentclass[12pt]{article}
\usepackage{epsfig}
\usepackage{amssymb}

\def\beq{\begin{equation}}
\def\eeq#1{\label{#1}\end{equation}}
\def\eeqn{\end{equation}}

\def\beqa{\begin{eqnarray}}
\def\eeqa#1{\label{#1}\end{eqnarray}}
\def\eeqan{\end{eqnarray}}

\let\bar=\overbar

\def\Dslash{\not{\hbox{\kern-4pt $D$}}}
\def\dslash{\not{\hbox{\kern-2pt $\del$}}}

\def\msb{{\bar{\ssstyle M \kern -1pt S}}}

\usepackage{fancyhdr,graphicx}
\def\Title#1{\begin{center} {\Large {\bf #1} } \end{center}}

\begin{document}

\Title{Deconfinement of quark content in neutron stars due to cosmic external agents}

\bigskip\bigskip

\begin{raggedright}

{\it M. \'Angeles P\'erez-Garc\'ia\index{Vader, D.}\\
Fundamental Physics Department \& IUFFyM\\
University of Salamanca \\
Plaza de la Merced s/n 37008 Salamanca, Spain\\
{\tt Email: mperezga@usal.es}}
\bigskip\bigskip
\end{raggedright}

\section{Introduction}

Whether or not the ultimate most dense state of matter is a quark-gluon plasma is uncertain to date. Since the pioneering work of Itoh \cite{itoh}  who predicted quark stars in the 1970's or that of Witten a decade later \cite{witten}, who pointed out that two-flavor $u-d$ matter would lower its energy by undergoing weak decay into $u-d-s$ quark matter, much work has been done. In our present knowledge of nature, the nucleon is a composite of quarks and gluons that are described by the Quantum Chromodynamics (QCD) theory  of strong interactions. In this framework  quarks have asymptotic freedom, and in systems where density is large enough they can be considered to move almost free and could hypothetically form an (extended) quark matter system \cite{collins}. At least compared to the intranucleon distances of $\sim 0.1$ fm$=10^{-16}$ m. 

Currently, our lack of some crucial experimental inputs to study the extended nuclear or quark matter systems prevent us from having a precise knowledge of what would be the characterization of the interior high density core in compact astrophysical systems,  such as neutron stars (NS) or, under the before-mentioned  possibility of quark deconfinement, the hypothesized quark stars. Since these are macroscopic many-body systems, nowadays not accessible to full simulation, one effective approach is to study the bulk thermodynamical properties of smaller size systems. In this way one can construct in a selected formalism, thermodynamical quantities, as i.e. pressure, $P$, or energy density,  $\epsilon$, to be subsequently used in the equation of state (EoS) of matter constituents, $P(\epsilon, T)$, at a given temperature, $T$. Considering a realistic and detailed description of matter is extremely important since, for example, pioneering works from Oppenheimer and Volkoff \cite{ov} found that a pure system of non-interacting neutrons could only provide Fermi  pressure support for a star with mass smaller that one solar mass, $\sim 0.7 M_\odot$, not accounting for objects with measured masses in excess of one solar mass.
\subsection{Neutron stars}
The astrophysical object arising after a stage of core collapse of a star with $M\sim (10-25) M_\odot$ is highly compact and degenerated. Typical compactness ratio or mass-to-radius ratio, $M/R$, can be obtained from a variety of means, namely  indirect measurements of masses and radii or even gravitational redshift $z\sim(1-R_S/R)^{-1/2}-1$ ($R_S$ is the Schwarzschild radius). The Fermi energies of the fermion populations in central regions, mostly neutrons, are typically $E_{\rm F}\simeq k^2_{\rm F}/2m\sim 60$ MeV at nuclear saturation density ,  $n_{0}\sim 0.17 fm^{-3}$, so that $n=k^3_{\rm F}/3\pi^2$, with $k_{\rm F}= 1.7\, fm^{-1}$ assuming {\it only one} fermionic component. We know that this is only approximately true since weak interaction must provide additional hadrons and leptons in the system conserving the baryon number. In addition, leptonic components are due in order to mantain charge neutrality. 

Since typical temperatures in the core collapse are at most of the order of $\sim40$ MeV, this is not enough to excite temperature fluctuations capable to deconfine the quark content in the nucleons with binding energies $\lesssim1$ GeV at typical densities of $n_{0}$ or in mass units $\rho_0\sim 2.4\times 10^{14}\, \rm g/cm^3$. Therefore these conditions are well within the degeneracy regime since $\xi =k_B T/E_F <<1$. This then poses a problem for possible mechanisms of effective quark deconfinement. For young objects, even the temperatures in the hot early ages of the collapse seem not high enough to drive the transition to a quark system. 

Let us recall that in the matter density-temperature diagram the quark hadron transition is supposed to happen at a temperature of the order of the QCD scale  $\Lambda_{QCD}\sim 170$ MeV for low chemical potential. In a dense medium the quark-hadron phase transition can happen at smaller temperatures. 
Therefore to excite the quark energies and be able to overcome the confining potential barrier one may think of an {\it external agent} and a mechanism that may trigger locally or globally this transition. One of the possibilities is a Majorana particle, with mass $m_X$, self- or co-annihilating to provide enough energy to the system, based on its massive character $E\sim m_Xc^2$ so that a percolation transition may happen. The nuclear system would be populated with energetically stable quark drops or nuggets of finite size as a result of steady reaction from a $X$ incoming flux. We discuss in the following sections about the feasibility and consequences of this scenario in the astrophysical context. 

\section{External cosmic agents}

One possibility to obtain the energy boost to trigger a percolation phase transition in the nuclear NS core could be provided by a (co-) self-annihilating particle. In the context of current cosmological problems dark matter (DM) provide several candidates that could fit into this picture. In particular, the lightest supersymmetric particle, the neutralino, seems to be a popular candidate in the overall phenomenological picture. Current experiments in direct detection based on nuclear recoils seem to point toward a ligth particle.  For example, the DAMA/LIBRA experiment \cite{dama} who find indications of $m_{X}\sim 4-12$ $\rm GeV/c^2 $ and seems to be confirmed by CoGEnt \cite{cogent}  and more recently some preliminar hints  from CDMS II \cite{cdmsII}, while there are other null-searches. In addition, indirect searches are based on observation of the galactic center or the sun. On this respect much attention has been paid to a possible line detected by Fermi \cite{fermi-line} from such processes although some issues must be understood in light of unexpected results \cite{halo-line}. Recent analysis by AMS-02 \cite{ams02} seem to confirm early measurements of a positron flux excess compatible with the annihilation of a particle with  $m_{X} \sim 0.4-1$ $\rm TeV/c^2$ \cite{hooper}. It remains to be seen if the positron flux suffers a dramatic decrease at higher energies since, in that case, could be signaling the existence of a DM self-annihilating reaction producing such an excess. Finally, accelerator searches in high-energy colliders, e. g. LEP and LHC, put also some constraint on DM properties in an effective field theory formalism by looking to the existence of mono-jet and mono-photon \cite{kopp}. They are more sensitive to low mass searches than current detectors based in nuclear recoils or indirect effects.
\subsection{Gravitational accretion by NSs}

This external cosmic agent could be taken as a DM component forming a galactic halo density profile. So far there is uncertainty about its actual shape,  but one popular option is that developed by Einasto \cite{einasto} which reads  
\begin{equation}
\rho_{X}(r)=\rho_{-2}e^{\frac{-2}{\alpha} [ (\frac{r}{r_{-2}})^{\alpha} -1]} ,
\end{equation}
with parameters $\rho_{-2}=0.22 \,\rm GeV/cm^3$, $\alpha=0.19$, and $r_{-2}=16 \,\rm kpc$ so that at the solar neighborhood the Keplerian velocity is $v \sim 220$ km/s and the local DM density is $\rho_{X, 0} \sim 0.8\, \rm GeV/cm^3$. 
These cosmic external particles are supposed to be thermalized in the galactic halo and therefore follow a Boltzmann distribution. Let us note that these particles in the $\sim 1\,\rm GeV/c^2-1\,TeV/c^2$ range, should have a non-relativistic quantum nature since $m_Xc^2/pc>> 1$. 

As the NS populates the galactic regions, $X$ agents may suffer a strong gravitational capture and be accreted onto it. Current indirect searches are based on this effect. As they scatter a few times internal dense matter they tend to drift to the center where they pile-up.

In this contribution we are interested in the possible implications of a gravitational accumulative effect  of these particles, in particular its effect on NSs.  If the considered external agent particle (co-) self-annihilates or decays in the central regions on these objects this could provide the energetic boost or spark needed to deconfine the quark content in the cold baryonic system. Same type of density enhancement effects are key to current mechanisms of production of an extra radiation component  in the sun \cite{dm-sun}.

Then, DM from this profile can be accreted by gravitational capture \cite{Gould} at a rate at peak of NS distribution ($\sim3$ kpc where $\frac{\rho_{\chi}}{\rho_{\chi, 0}}\sim 30$) for a typical NS with $R\sim 12$ km and $M\sim 1.5M_\odot$,
\begin{equation}
 {\cal C} \approx \frac{2.7 \times 10^{27}}{ m_{\chi} (\rm GeV)}\, \mathrm{particles\, s^{-1}}\, ,
\end{equation}
but off-peak, at about solar circle it may be reduced somewhat. In this estimate, we assume a WIMP-nucleon (spin independent) cross section $\sigma=\sigma_{\chi N}=7\,10^{-41}\,\rm cm^2$ \cite{bertone}. Once the steady state, resulting from competing processes of annihilation and accretion,  has been reached, the elapsed time is $\tau_\mathrm{DM} \approx \frac{1}{\sqrt{ {\cal C}<\sigma_a v>/V}}$ where $V$ is the volume of the star. Typically, if we assume a light  $\sim 10$ $\rm GeV/c^2$ DM particle, as direct detections experiments seem to preliminary suggest, we obtain $\tau_\mathrm{DM}\sim 3.5 \,\times\, 10^3$ yr. For a velocity (temperature) dependent cross section and heavier DM particles $\sim 100$ $\rm GeV/c^2$, this delay can be longer, $\tau_\mathrm{DM} \ge 6 \,\times\,10^5$ yr  \cite{kou08}. We have taken $<\sigma v >\simeq 3 \times 10^{-26} \,\rm cm^3 s^{-1}$. With this simplification, the population of agents at time $t$ is given by
\begin{equation}
N(t)= {\cal C} \tau\,{\rm tanh}(t/ \tau),
\end{equation}
The distribution radius in this region  depends on the nature of the DM particle being accreted.  Assuming a regime when velocities and positions of the X's  follow a Maxwell-Boltzmann distribution with respect to the centre of the NS, the thermalization volume in the compact star has a radius
\begin{equation}
r_{\rm th}= \left(\frac{9 k_B T_c}{4 \pi G \rho_c m_{\chi}}\right)^{1/2}=20.2\, \left(\frac{T_{\rm c}}{10^{\rm 7} K}\right)^{1/2} \left(\frac{10^{\rm 14} {\rm g/cm^3}}{\rho_{\rm c}}\right)^{1/2} \left(\frac{10 \,{\rm GeV/c^{\rm 2}}}{m_{\chi}}\right )^{1/2}\,(\rm m).
\end{equation}
This volume is just a tiny value of the star core, $V_X\sim 5\times 10^{-9}V_0$ but it could drive dramatic consequences since this energy is vastly injected in this region.
 
\section{Deconfinement}
It is in this steady state of DM accretion that the possibility of self-annihilation is larger, since it grows with density $n_X$ squared. Then the luminosity at a given volume can be obtained by integration as
\begin{equation}
L= <\sigma v>{m_X}c^2\int_0^{r_{\rm th}} n^2_X(r) d^3\bf r.
\end{equation}
The spectrum of the energy deposited is so far unknown but a simple energy count allows to produce a fraction of energy $E\sim m_Xc^2$. From a given threshold, $E_0$, the energy produced in photons in an interval $\Delta_E=E-E_0$ is given by
\begin{equation}
L_{\Delta_E}= \frac{<\sigma v>}{2} \left[\int_{E_0}^{E} E' \frac{dN_{ph}}{dE'} dE'\right] \int_0^{r_{\rm th}} n^2_x(r) d^3{\bf r}.
\end{equation}
where $\frac{dN_{ph}}{dE}$ is the spectrum of  photons produced in a single annihilation event and depends on the specific nature of the X agent.

Typical masses for the external cosmic agents are in the range $m_X\sim1$ $\rm GeV/c^2$ to 1 $\rm TeV/c^2$. They could provide enough energy per baryon to deconfine the quark content in typical NS interior conditions if the fraction of deposited energy,  $f$, is large. The energy rate deposited in the medium is estimated in a steady state as 
\begin{equation}
{\dot E_{\rm annih}} ={\cal C} f m_{X}c^2 ,
\label{edot}
\end{equation}
where $f \approx 0.01-1$. Let us note here that typically there are suppressed channels to $\cal O (\alpha)$  such as the internal electron bremsstrahlung but the $\pi^0$ decay at tree level seems to be more extended to lower energies. In particular highly energetic neutrinos would be trapped by matter so that $f$ would be close to unity. 

\begin{figure}[htb]
\begin{center}
\epsfig{file=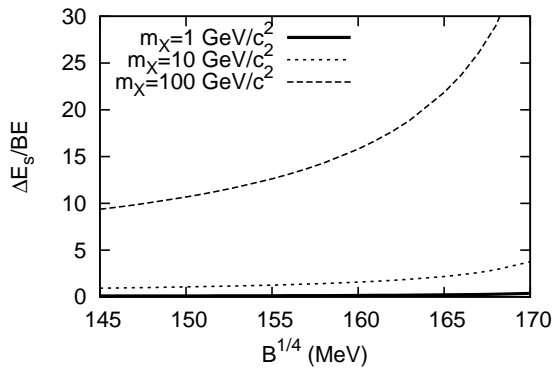,scale=0.6, angle=-90}
\caption{Ratio of spark energy over the strangelet binding energy with a deposition fraction $f=0.1$ . We have assumed an $A=10$ quark nugget with density $2n_{A,0}$.}
\label{fig:spark}
\end{center}
\end{figure}

In Fig~\ref{fig:spark} \cite{mor} we show the fraction of released energy $\Delta E_S$ in one process involving an efficiency $f=0.1$ over the binding energy (BE) of a stable non-decaying nugget of deconfined quark matter as a function of the MIT bag constant, $B$, assuming reasonable values in the quark interaction \cite{mit}.  We have considered a nugget with $A=10$ which is the minimum value for stability \cite{amin}, with internal density of $n_A=2n_{0}=2n_{A,0}$. We have considered $X$-masses of $1, 10, 100$ $\rm GeV/c^2$ with solid, short dashed, long dashed lines respectively. For $X$ agents with masses above a few $\rm GeV/c^2$ the energy release could be up to an order or magnitude larger than the typical $\lesssim$ GeV quark binding energy in nucleons allowing for the deconfinement. Let us recall that estimations are $\Lambda_{QCD}\sim {\cal O}(10 \,\rm MeV)$ at large densities, and therefore the deconfinement radius could grow leading to an asymptotic deconfinement \cite{ri}.

\begin{figure}[htb]
\begin{center}
\epsfig{file=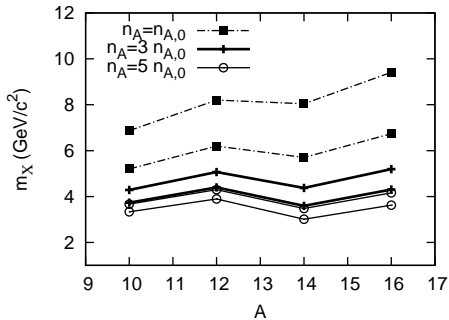, angle=0, scale=1}
\caption{Values of $X$ mass producing deconfinement as a function of $A$ for strangelet with different densities and $f=0.9$. Each curve shows limiting values of the bag constant, $B$.}
\label{fig:mass}
\end{center}
\end{figure}

In Fig~\ref{fig:mass} we show for different values of the central density inside the nugget,  $n/n_{A,0}= 1, 3, 5$, and for limiting values of the bag constant, allowed values of $X$ mass with $f=0.9$  producing strangelets with baryonic number $A<17$ \cite{perez10}. Let us notice that although it is uncertain there should be a quark nugget chart of stability, similar to the one we know on the more familiar non-strange nuclei.

As seen, the physical picture points to a sort of {\it Trojan mechanism} where the larger and more dense stars can  accrete this extra component that could trigger a spark seeding that may lead to a depletion of the internal pressure and further consequences. 
The possibility that the nuclear equation of state (EoS) undergoes a phase transition to a quark system has been considered  already in the literature, for a review see \cite{eos} \cite{ouyed}. Based on energetic stability criteria, the $u-d$ quark and a subsequent weak decay populating the $s$ quark fraction causes a decrease of the energy density as a function of density. In this sense the light quark sector is the most relevant since masses are at the reach of the 
energy range considered. In addition, lepton populations are included in order to achieve electrical charge neutrality.
\subsection{Emission of Gamma ray bursts}

It is thought that the {\it macroscopic} conversion of the full or hybrid star could happen very rapidly, in less than a second \cite{horvath10}. This event may produce, as a consequence, an effect  on the kinematics with a birth kick and rotation as a result of the partial burning of the NS \cite{perez12},  being accompanied by the emission of a short gamma ray burst (GRB). In particular it is expected that  the duration of the {\it very short} GRBs is less than $\sim 0.1$ s. These type of GRBs show a larger spectral hardness ratio, as compared to longer GRBs and the isotrotropic equivalent energies involved are of the order of $E_{\rm iso}\sim 10^{48}-10^{52}$ erg. It has been seen that short GRBs are mostly isotropically distributed but not for the very short GRBs \cite{cline} and tend to have small redshifts $z<2$.

\begin{figure}[htb]
\begin{center}
\epsfig{file=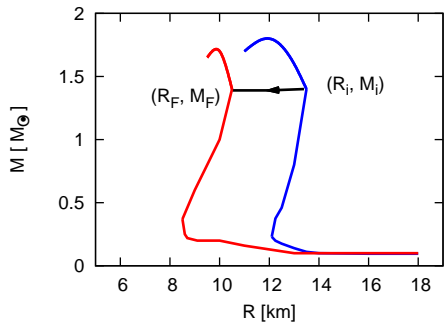, angle=-90, scale=0.5}
\caption{Mass-radius diagram showing the nuclear and (hybrid) quark configurations from a static approach}
\label{fig:mr}
\end{center}
\end{figure}

If this deconfinement transition takes place, the energies emitted  should be such that there would be a migration of the initial compact object in the Mass-radius diagram following a NS trajectory into the allowed configurations over the quark EoS. In Fig~\ref{fig:mr} \cite{perez13} we show schematically the conversion process where one NS configuration (blue curve) with initial mass and radius $(M_{\rm i}, R_{\rm i})$ suffers an internal  transition to a quark final configuration characterized by $(M_{\rm F}, R_{\rm F})$ (red curve).
 
It has been seen that the emission of the very powerful gamma ray burst is mostly due to the gravitational and internal energies being released \cite{bombaci}. The quark deconfinement allows the configuration of more compact objects with a radius in the $R\sim 7-12$ km range. It is therefore expected that the release is about $E_{grav}\sim E_{internal}\sim 10^{53}\frac{\Delta R}{R}$ according to Ma et al. \cite{ma}.

This type of conversion is thought to be accompanied by a multimessenger emission of the before-mentioned gamma ray burst, along with gravitational waves and possibly cosmic rays \cite{kotera}. If most of the gamma-rays are produced at the photosphere, one should expect a thermal (quasi-Planckian) spectrum. The peak energy of the observed spectrum will then be located at $E_\mathrm{p}\simeq 3.9\,
k_\mathrm{B} T_\mathrm{ph}$, where $T_\mathrm{ph}$ is the temperature of the photosphere. It can be computed assuming an adiabatic radial expansion from the ejection to the photosphere \cite{perez13}:
\begin{equation}
E_\mathrm{p}\simeq 3.92 \left(\frac{3 E}{16\pi a R_\mathrm{NS}^2 \Delta}\right)^{1/4} \left(\frac{R_\mathrm{ph}}{R_\mathrm{sat}}\right)^{-2/3}\simeq 
18\, M_\mathrm{ej,-5}^{-1} f_\mathrm{ej,-3}^{11/12} \Delta t_{-6}^{-1/4}\, \mathrm{keV}\, ,
\end{equation} 
where $a$ is the Stefan-Boltzmann constant and  $\Delta = c \Delta t$ is the  outer crust width and more specifically $\Delta t=\Delta t_{-6} \times 10^{-6}\, \mathrm{s}$ is the light crossing time ($\Delta =\Delta t_{-6}\times 300\, \mathrm{m}$). $M_\mathrm{ej}$ is the ejected mass in solar mass units. To be consistent in the production of gamma rays one needs masses $M_\mathrm{ej}\lesssim 10^{-4} M_\odot$ \cite{perez13}. The fraction of energy devoted to eject the external crust in the event is labeled as $f_\mathrm{ej}$. We use the notation usual in astrophysical arguments here.

The saturation radius $R_\mathrm{sat}$ refers to the radial value where the Lorentz factor $\Gamma$ of debris is saturated to a maximum value after the initial expansion
 \begin{equation}
R_\mathrm{sat} \simeq
\Gamma 
R \simeq 2
\times 10^7\, M_\mathrm{ej,-5} ^{-1} f_\mathrm{ej,-3}\, \mathrm{cm}\, ,
\end{equation}
assuming an initial radius $R_\mathrm{ej}=R-\Delta\simeq R_\mathrm{NS}=12$ km for the ejection.  The ejecta will become transparent for its own radiation at the photospheric radius
\begin{equation}
R_\mathrm{ph} \simeq \sqrt{\frac{\kappa M_\mathrm{ej}}{4\pi}} \simeq 2\times 10^{13}\, M_\mathrm{ej,-5}^{1/2}\, \mathrm{cm}\, ,
\end{equation}
where we take the Thomson opacity $\kappa\simeq
\, 0.2 \,\rm cm^2/g$. Here we assume $Y_\mathrm{e}=0.5$ free electron per nucleon in the expanding gas. 

The efficiency of the photospheric emission can also be deduced from the adiabatic evolution and equals $f_\mathrm{\gamma}\simeq \left(R_\mathrm{ph}/R_\mathrm{sat}\right)^{-2/3}$. This leads to a better estimate of the isotropic equivalent energy radiated in gamma-rays,
\begin{equation}
E_\mathrm{\gamma,iso} \simeq 2\times 10^{48}\, \left(\frac{f_\mathrm{b}}{50}\right) M_\mathrm{ej,-5}^{-1} f_\mathrm{ej,-3}^{5/3}\, \mathrm{erg/s}\, ,
\end{equation}
that stands on the lower values for GRBs. This energy includes a factor due to beaming as is emitted by a relativistic outflow  within an opening angle $\theta_\mathrm{j}$ so that the true energy release is $E_\mathrm{\gamma}=f_\mathrm{b}^{-1} E_\mathrm{\gamma,iso}$, where the beaming factor is defined as
\begin{equation}
f_\mathrm{b}=\left(\frac{\Omega}{4\pi}\right)^{-1}=\left(1-\cos{\theta_\mathrm{j}}\right)^{-1}\simeq \frac{2}{\theta_\mathrm{j}^2}\,\, \, \, \, \,  \mathrm{if}\, \theta_\mathrm{j}\ll 1\, .
\end{equation}
Unfortunately, the opening angle $\theta_\mathrm{j}$ cannot be easily measured in GRBs but they are estimated in the $\theta_\mathrm{j}\simeq (1-30)^\circ$. This, of course, adds still a large uncertainty due to the unknown beaming factor $f_\mathrm{b}$, $ f_\mathrm{ej}$ and $M_\mathrm{ej}$ but nevertheless confirm the capacity to produce bright and hard spikes of gamma-rays in the considered scenario.

In conclusion we have presented a scenario where the existence of cosmic external agents able to deposit energy in the medium by (co-) or self-annihilation may drive a phase transition. This would have macroscopic consequences, by changing the EoS of the object and by the emission of a multimessenger signal, than in the photon range could be detected with gamma rays of $\sim10-100$ keV.

\section{Acknowledgements}
I am grateful to the LOC of this meeting for the invitation to participate and  financial support received. We additionally thank the Spanish Ministry of Science and Innovation  MULTIDARK, FIS2011-14759-E and FIS2012-30926 projects for partial financial support.


\begin{thebibliography}{99}

%

\bibitem{itoh} N. Itoh,  Progress of Theoretical Physics, {\bf 44}, 291 (1970)
\bibitem{witten} E. Witten, Phys. Rev. D,{\bf 30}, 272 (1984)
\bibitem{collins} J. C. Collins and M. J. Perry, Phys. Rev. Lett. {\bf 34}, 1353
(1975)
\bibitem{ov} J. R. Oppenheimer and G. M. Volkoff, Phys. Rev. {\bf 55}, 374, (1939) 
\bibitem{dama} R. Bernabei, P. Belli, F. Cappella et al., Eur. Phys. J. C {\bf 67}, 39 (2010)
\bibitem{cogent} C. E. Aalseth et al. [CoGeNT Collaboration], Phys. Rev. Lett. {\bf 106}, 131301, (2011)
\bibitem{cdmsII} R. Agnese, Z. Ahmed, A.J. Anderson et al, arXiv:1304.4279 
\bibitem{fermi-line} T. Bringmann, X. Huang, A. Ibarra, S. Vogl and C. Weniger, arXiv:1203.1312
\bibitem{halo-line} D. Whiteson, arXiv:1302.0427
\bibitem{ams02}  M. Aguilar et al, Phys. Rev. Lett. {\bf 110}, 141102, (2013)
\bibitem{hooper}  I. Cholis, D. Hooper, arXiv:1304.1840; J. Kopp, arXiv:1304.1184 
\bibitem{kopp}   J. Kopp, arXiv:1105.3248
\bibitem{einasto} J. Einasto, Trudy Inst. Astroz. Alma-Ata, {\bf  17}, 1, (1965),
J. Einasto, Astronomische Nachrichten, {\bf 291}, 97, (1969)
\bibitem{dm-sun} S. Adri\'an-Martinez, I. Al Samarai, A. Albert et al, arXiv:1302.6516
\bibitem{Gould} A. Gould, ApJ, {\bf 321}, 571 (1987)
\bibitem{bertone} G. Bertone, ed., Particle Dark Matter: Observations, Models and Searches. Cambridge University Press, (2010), ISBN 978-0-521-76368-4
\bibitem{kou08} C. Kouvaris, Phys. Rev. D {\bf 77}, 023006 (2008)
\bibitem{mor} M. A. Perez-Garcia, the Proceedings of the Rencontres de Moriond Cosmology 2012, La Thuile, Aosta valley, Italy, arXiv:1205.2581v1 
\bibitem{mit} A. Chodos et al. 1974, Phys Rev. D {\bf 10}, 2599, (1974)
\bibitem{amin} J. Madsen, Phys. Rev. Lett. {\bf 87}, 172003, (2001)
\bibitem{ri} D.H. Rischke and M.A. Stephanov, Phys Rev Lett., {\bf 87}, 062001 (2001)

\bibitem{perez10}  M.~A. Perez-Garcia,  J. Silk and  J. R. Stone,  Physical Review Letters, {\bf 105}, 141101 ,(2010)
\bibitem{eos} N. K. Glendenning, {\it Compact stars}, Ed. Springer-Verlag, New York (2000)
\bibitem{ouyed} R. Ouyed, J. Dey and M. Dey, arXiv:astro-ph/0105109
\bibitem{horvath10} J. E. Horvath,  arXiv:1005.4302
\bibitem{perez12} M. A. Perez-Garcia, J. Silk,  Phys. Lett. B, {\bf 711},  6, (2012)

\bibitem{cline}  D.~B. Cline, C. Matthey, and S. Otwinowski ApJ, {\bf 527}, 827 (1999)

\bibitem{perez13} M. A. Perez-Garcia, F. Daigne and J. Silk, {\bf 768},  ApJ, 145 (2013) 

\bibitem{bombaci}I.  Bombaci, I. Parenti, I. Vida\~na, ApJ,  {\bf 614}, (2004) 

\bibitem{ma} F. Ma and B. Xie, ApJ, {\bf 462}, L63 (1996)

\bibitem{kotera} K. Kotera, M. A. Perez-Garcia and J. Silk, arXiv:1303.1186 


\end{thebibliography}
\end{document}